\documentclass[]{spie}  

\usepackage{amsmath,amsfonts,amssymb}
\usepackage{graphicx}
\usepackage[colorlinks=true, allcolors=blue]{hyperref}

\usepackage{graphicx}
\usepackage{subcaption}
\usepackage[export]{adjustbox}
\usepackage{wrapfig}

\title{The Power Supply Unit onboard the HERMES nano-satellite constellation}

\author[a]{Paolo~Nogara}
\author[a]{Giuseppe~Sottile}
\author[a]{Francesco~Russo}
\author[a]{Giovanni~La~Rosa}
\author[a]{Fabio Paolo~Lo~Gerfo}
\author[a]{Melania~Del~Santo}
\author[b]{Yuri~Evangelista}
\author[c]{Riccardo~Campana}
\author[c]{Fabio~Fuschino}
\author[d]{Fabrizio~Fiore}

\affil[a]{INAF/IASF-Palermo, Via Ugo la Malfa, 153, 90146 Palermo PA, Italy}
\affil[b]{INAF/IAPS, Via del Fosso del Cavaliere, 100 - 00133 Roma, Italy}
\affil[c]{INAF/OAS, Via Piero Gobetti, 93/3, 40129 Bologna BO, Italy}
\affil[d]{INAF/OATS, via G.B. Tiepolo, 11, 34143 Trieste, Italy}

\authorinfo{Further author information: (Send correspondence to Paolo Nogara and Giuseppe Sottile)\\ Paolo Nogara: E-mail: \url{paolo.nogara@inaf.it}, Telephone: +39 091 6869 567\\ Giuseppe Sottile: E-mail: \url{giuseppe.sottile@inaf.it}, Telephone: +39 091 6869 567}

\begin{document} 
\maketitle

\begin{abstract}
HERMES Pathfinder (High Energy Rapid Modular Ensemble of Satellites Pathfinder) is a space mission based on a constellation of nano-satellites in a low Earth Orbit, hosting new miniaturized detectors to probe the X-ray temporal emission of bright high-energy transients such as Gamma-Ray Bursts and the electromagnetic counterparts of Gravitational Waves. This ambitious goal will be achieved exploiting at most Commercial off-the-shelf components. For HERMES-SP, a custom Power Supply Unit board has been designed to supply the needed voltages to the payload and, at the same time, protecting it from Latch-Up events.
\end{abstract}


\section{INTRODUCTION}
\label{sec:intro}  


HERMES (High Energy Rapid Modular Ensemble of Satellites)\cite{Fiore2} is a mission based on a constellation of nano-satellites (3U CubeSats) in low Earth orbit (LEO), hosting innovative X and $\gamma$-ray detectors to probe the high-energy emission of bright transients\cite{Sanna1,Burd3,Evang4}. HERMES is built upon a twin project: the HERMES Technological Pathfinder (HERMES-TP), 
and the HERMES Scientific Pathfinder (HERMES-SP).

Both projects (HERMES-TP and HERMES-SP) provide three complete satellites (payload and service module) to the constellation, with the aim of demonstrating that fast Gamma-Ray Burst (GRB) detection and localization is feasible by disruptive technologies on-board miniaturized spacecrafts, mostly exploiting commercial off-the shelf (COTS) components at a cost 1–2 orders of magnitude lower than space qualified ones (e.g. ESA M-class missions and NASA Explorer missions) and with a development time of only a few years.

Being based on cost-effective nano-satellites, HERMES is intrinsically a modular project. This allows to avoid single or even multiple-point failures (if one or several units are lost, the constellation and the experiment is guaranteed) and to test the in-flight hardware and the on-board software since the first launches. If needed, both hardware and software can be improved with the following launches.

 In this paper first we will give a brief description of the payload, then we will describe the custom Power Supply Unit (PSU) implemented and the design choices that we made. Finally, we will present the different components, with their peculiarities, that make up the PSU and a few results of our tests. \\

\section{The Payload}
\begin{figure} [h!]
   \begin{center}
   \includegraphics[height=12cm]{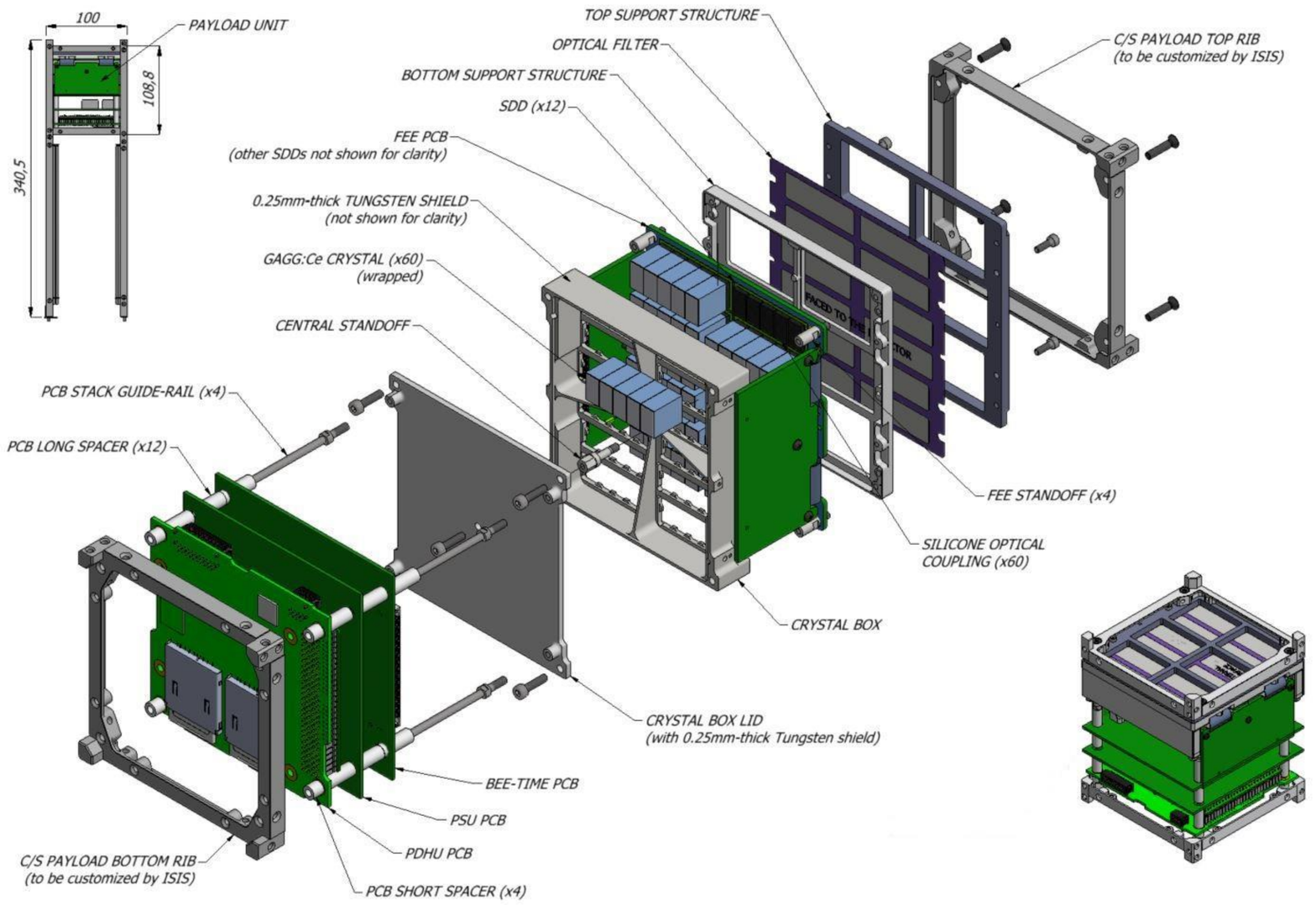}
   \end{center}
   \caption{Exploded view of the HERMES Payload (P/L).}
  \label{fig:figu1} 
   \end{figure}

\begin{figure} [ht!]
   \begin{center}
   \includegraphics[height=20cm]{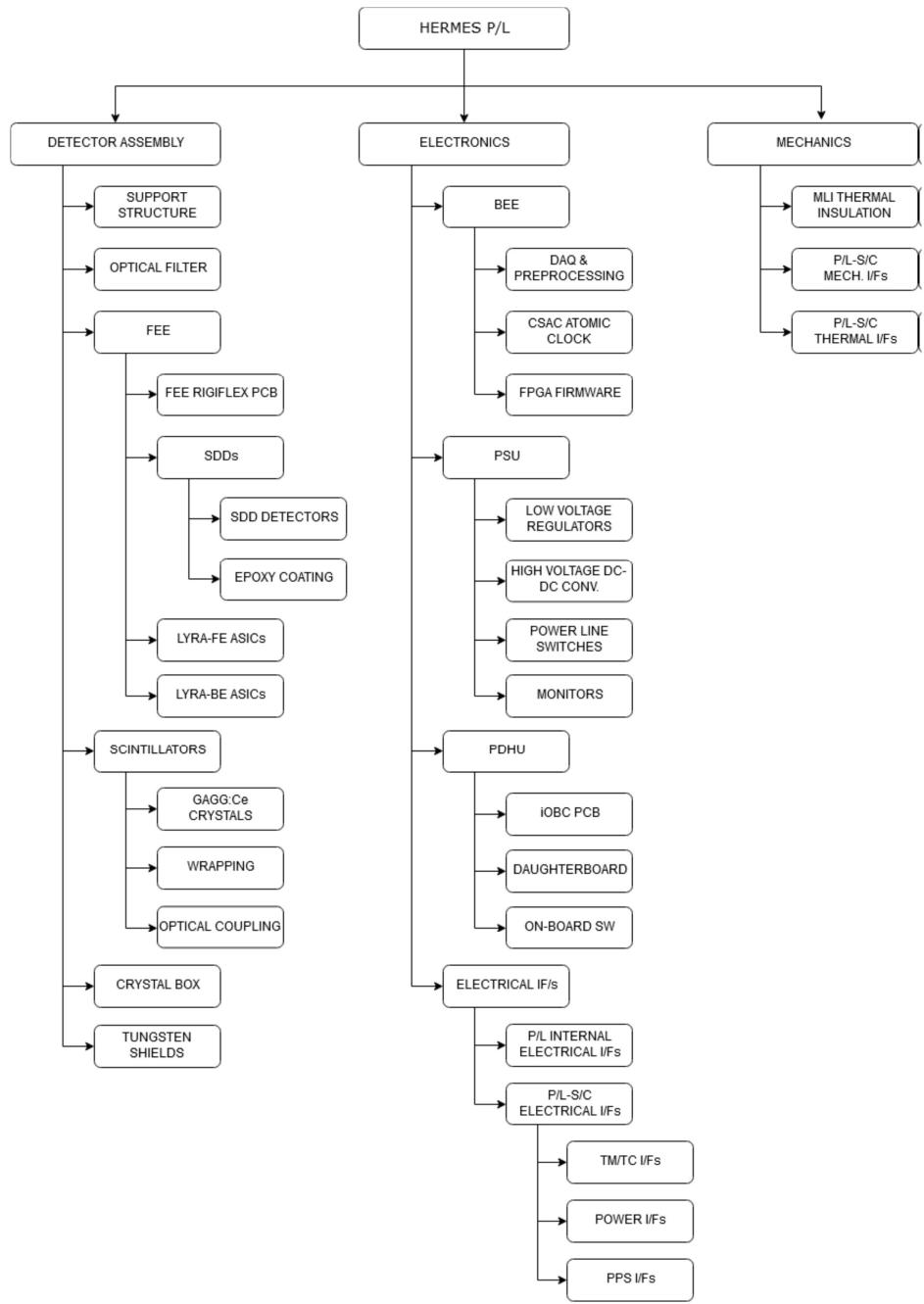}
   \end{center}
   \caption{HERMES P/L Product tree.}
  \label{fig:figu2} 
   \end{figure}

The HERMES Payload (P/L\cite{Evang4}) is hosted in a single CubeSat Unit (1U) and it is composed by a number of sub-systems, as represented in Figure \ref{fig:figu1} and in the P/L product tree (Fig. \ref{fig:figu2}).
In particular:
\begin{itemize}
    \item \textbf{Detector assembly (DA)}
        \begin{itemize}
        \item \textbf{Detector support structure}, mounted on the top of the detectors plane. This structure is split in an upper and a lower part with the optical filter in-between. The lower detector support structure is shaped in order to avoid touching the electrical components on the Front End Electronics (FEE);
        \item \textbf{Optical filter}, providing effective filtering of UV-VIS-IR bright emission from the Earth;
        \item \textbf{FEE} board, which includes:
        \begin{itemize}
            \item \textbf{Silicon Drift Detectors (SDDs)} mounted on the printed circuit board (PCB) side opposite to the detector support structure;
            \item the first analogue stage (LYRA-FE) of the electronics for each channels (pre-amplifier, first shaper stage, line buffer);
            \item the power distribution and filtering for SDDs power;
            \item the second (mixed signal) stage of the electronics (LYRA-BE).
        \end{itemize}
        \item Scintillator crystals optically coupled to the SDD detectors which are contained in a light-tight \textbf{crystal box}.
    \end{itemize}
    \item \textbf{Back End Electronic (BEE)} board: is a PC/104 format board developed by INAF specifically for the HERMES payload, using COTS components. Within the payload section, the BEE board is placed between the Power Supply Unit (PSU), to which is directly plugged by two connectors, and the Detector Assembly.
    The main component of the BEE is the Intel Cyclone V E FPGA, which is optimized for applications that require low power consumption and low cost. The board hosts two identical quad-serial configuration (EPCQ) devices, that are an in-system programmable NOR flash memory. The redundant devices and the hardwired logic present on board guarantee the correct programming of the FPGA even if one of the two EPCQs has the configuration file corrupted. The BEE manages configuration tables of the LYRA ASICs through serial buses and monitors the trigger outputs to start the acquisition of the events by means of four ADAQ7980 16 bits 1MSPS  ADCs, one for each quadrant of the instrument.
    The time-tagging for the scientific events is generated by the combination of a GPS synchronization signal provided by the nanosatellite platform and the Microsemi SA.45s CSAC Chip Scale Atomic Clock (CSAC) hosted in the BEE board. The BEE manages all the power switches for the FEE and the housekeeping 8 bits ADC located on the PSU board. The Payload Data Handling Unit and BEE communicate by a reliable custom protocol based on SPI serial bus;
    \item \textbf{Power Supply Unit (PSU)} board, like the BEE board, is developed by INAF specifically for the HERMES payload using COTS components. PSU is connected to the satellite power bus, which is in charge of the generation, filtering and distribution of all the needed voltages to the payload.
    
    \textbf{FEE-BEE and FEE-PSU Harness} connect the FEE to the BEE and PSU, respectively. More details about the PSU board is described in the following sections.
    \item \textbf{Payload Data Handling Unit (PDHU)}, which configures the FEE and BEE, receives digital data from BEE and formats the event information (energy, address, time), generates the photon list, elaborates the data with a burst search algorithm, produces TM data packets, receives and manages telecommands (TLCs). The PDHU issues commands for the Power Supply Unit (PSU) to generate and distribute the required voltages levels for the BEE, FEE and SDD, and procures a safe switch on/off procedure when ramping up/down the relevant voltages.
    \item \textbf{Mechanical structure}.
\end{itemize}

   \clearpage

\section{Power Supply Unit board}
\label{sec:title}

A custom PSU board has been designed and manufactured to provide the
power supplies required by the payload. All the low voltages needed for the operation of the FEE are generated by ultra-low noise linear Low Drop-Out (LDO) regulators \cite{Fab-LDO}. A DC-DC converter (Picoelectronics 12SAR250) is used for the generation of the high-voltage detector biases, While a Texas Instrument LMZ30602 is in charge of the generation of the 1.1V required by the BEE FPGA core. 

The PSU must handle the switching on/off of the voltages required to supply the payload. These voltages come from the satellite BUS (SBUS) and must be properly supplied to the various sections of the Payload. 

The implementation of the various PSU sections was based on the fact that the PSU will have to operate in space. This gives strong constraints to the choice of the architecture of the different sections and of the components that make up the PSU. In addition, the choice to use commercial components was driven by the requirement to avoid high costs and long procurement times, such as in the case of the Radiation Hardened components.
 
   \begin{figure} [h!]
   \begin{center}
   \includegraphics[height=12cm]{{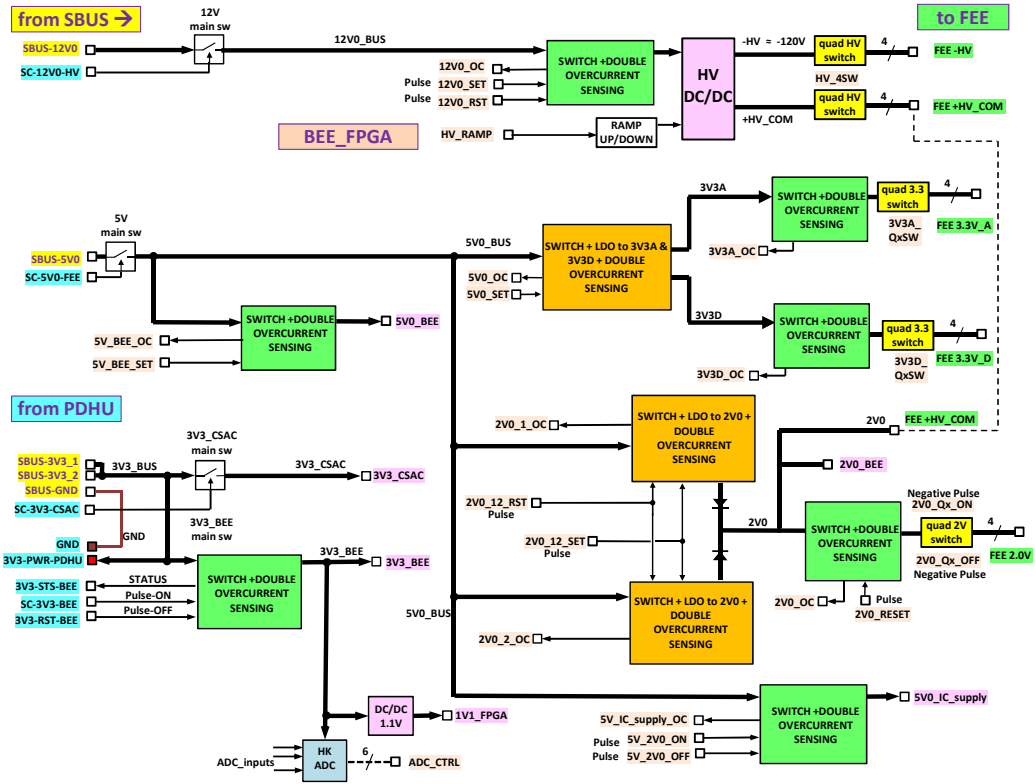}}
   \end{center}
   \caption{Block diagram of the HERMES PSU board.The bold lines represent the power rails while the thin ones depict the signals controlled by PDHU or BEE.}
  \label{fig:figu3} 
   \end{figure}

\begin{figure} [h!]
   \begin{center}
   \includegraphics[height=8cm]{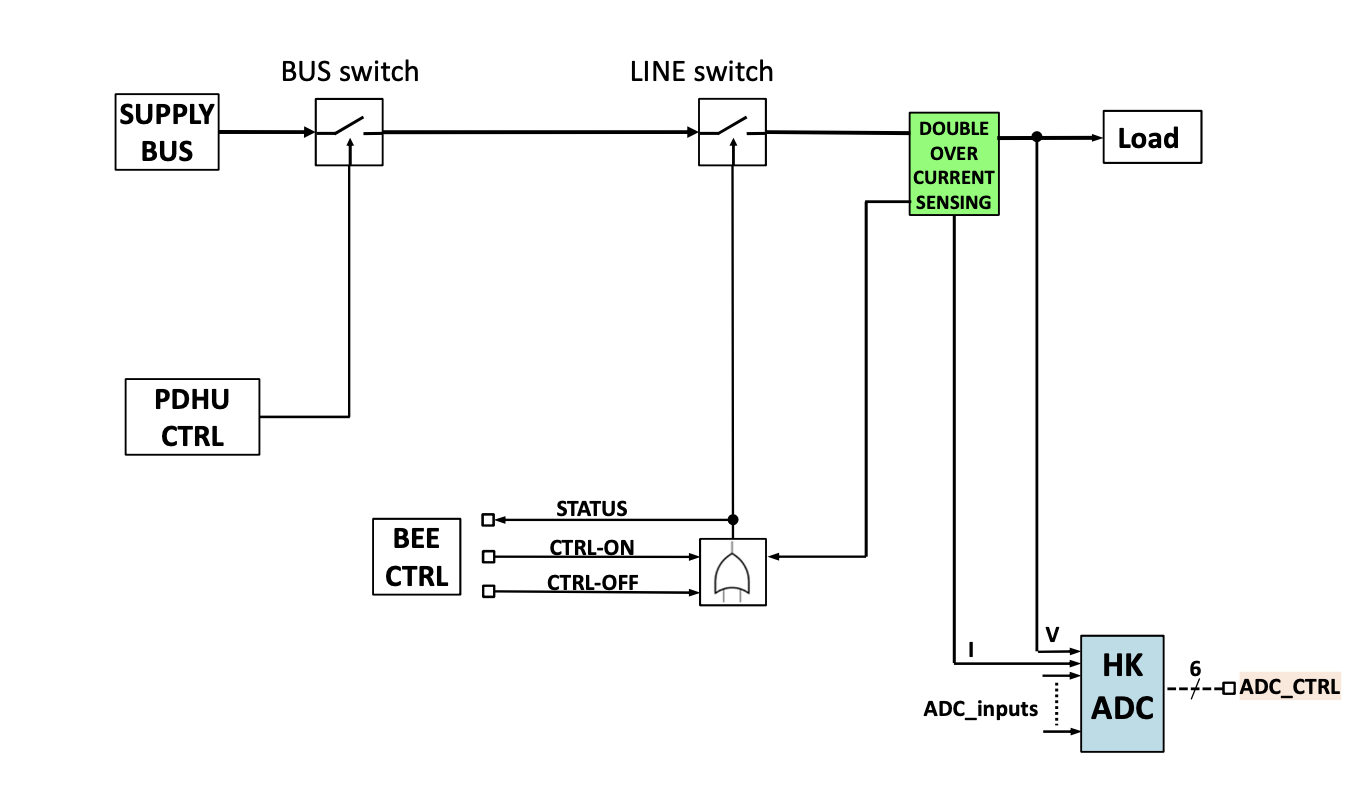}
   \end{center}
    	\caption{Block diagram of the general HERMES-PSU line.}
	\label{fig:figu4}
    \end{figure}

The architecture of the PSU board is sketched in Figure \ref{fig:figu3}.
The block diagram shows the differences between the power rails and the control lines driven by PDHU or by BEE accordingly of the supplied load. The SBUS lines provide three BUS voltages: 
\begin{itemize}
    \item \textbf{12V0} This power rail, controlled by BEE, provides the voltage to section that generates HV needed for SDD detectors;
    \item \textbf{5V0} This power rail provides voltage to several loads such as the LDOs, that supply the FEE, and the feedback network of the HV section.  It also feeds some BEE sections such as the four ADCs for the acquisition of the scientific events and the pulse generator for calibration of the detectors;
    \item \textbf{3V3} This power rail supplies the BEE and the ADC dedicated to the housekeeping data.
\end{itemize}
For each main power rail shown in Figure \ref{fig:figu3}, the same generic circuitry structure shown in Figure \ref{fig:figu4} is applicable.

The PDHU controls the BUS switch, to provide or remove power to the input of “Line switch”. This feeds the load and is driven by the OR logic of the BEE and the Overcurrent Sensing block. Generally, the BEE provides power to the load accordingly with the operative mode of the payload. The Overcurrent Sensing circuit monitors the current flowing on the line and quickly opens the Line switch if an overcurrent condition is detected, in order to prevent a damage of the load. The voltage and the current are digitally converted and constantly monitored by the BEE to evaluate the health status of the load. In most of the circuit architectures used on the PSU board, the Overcurrent Sensing block is supplied by the same voltage which is being monitored. This particular design guarantees the protection of the Overcurrent Sensing block itself from destructive latch-up (LU) events.

\subsection{3V3 section}
\begin{figure} [h!]
   \begin{center}
   \includegraphics[height=6cm]{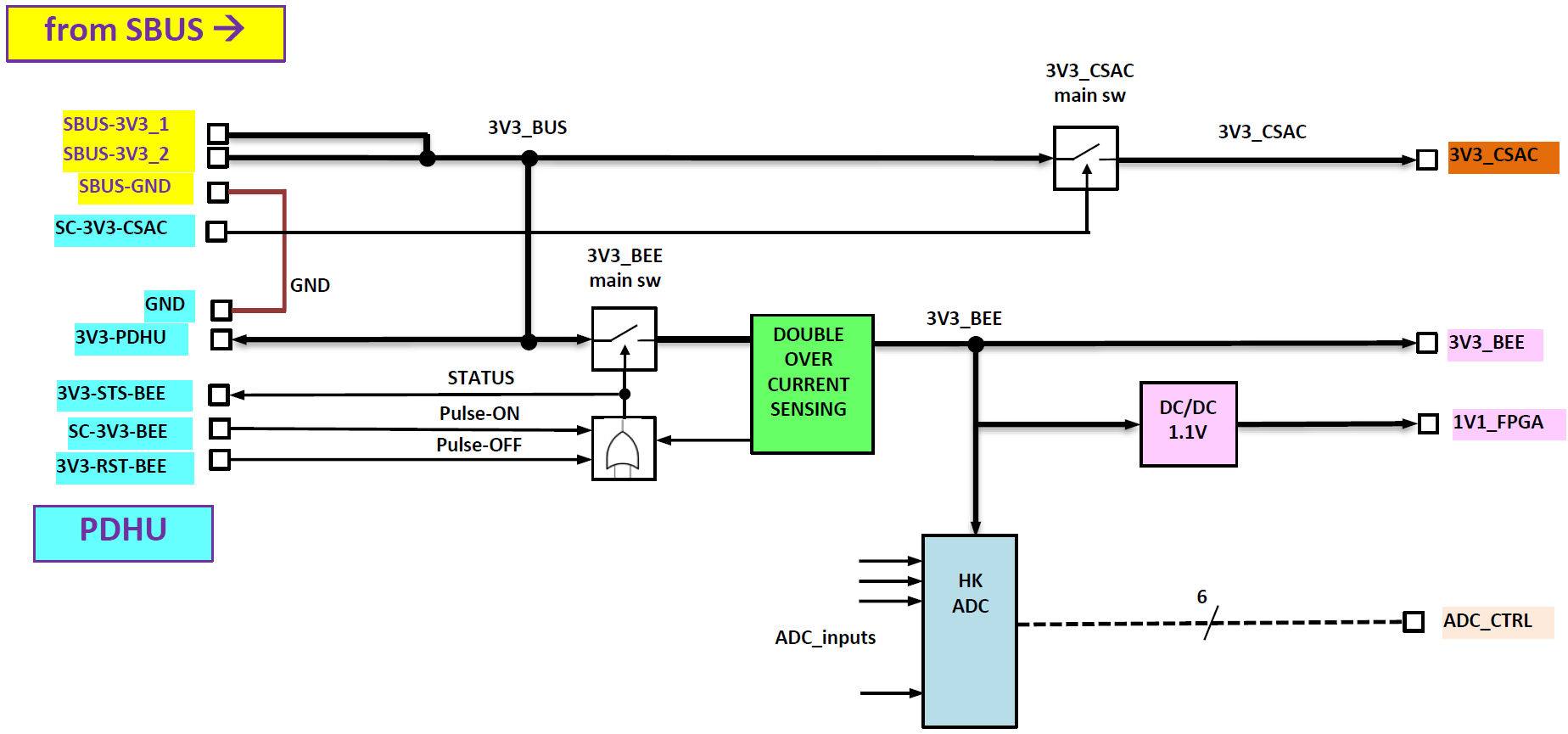}
   \end{center}
   \caption{Block diagram representation on the  of HERMES-PSU 3V3 section.}
  \label{fig:figu5} 

\begin{subfigure}{0.5\textwidth}
\includegraphics[width=0.9\linewidth, height=4cm]{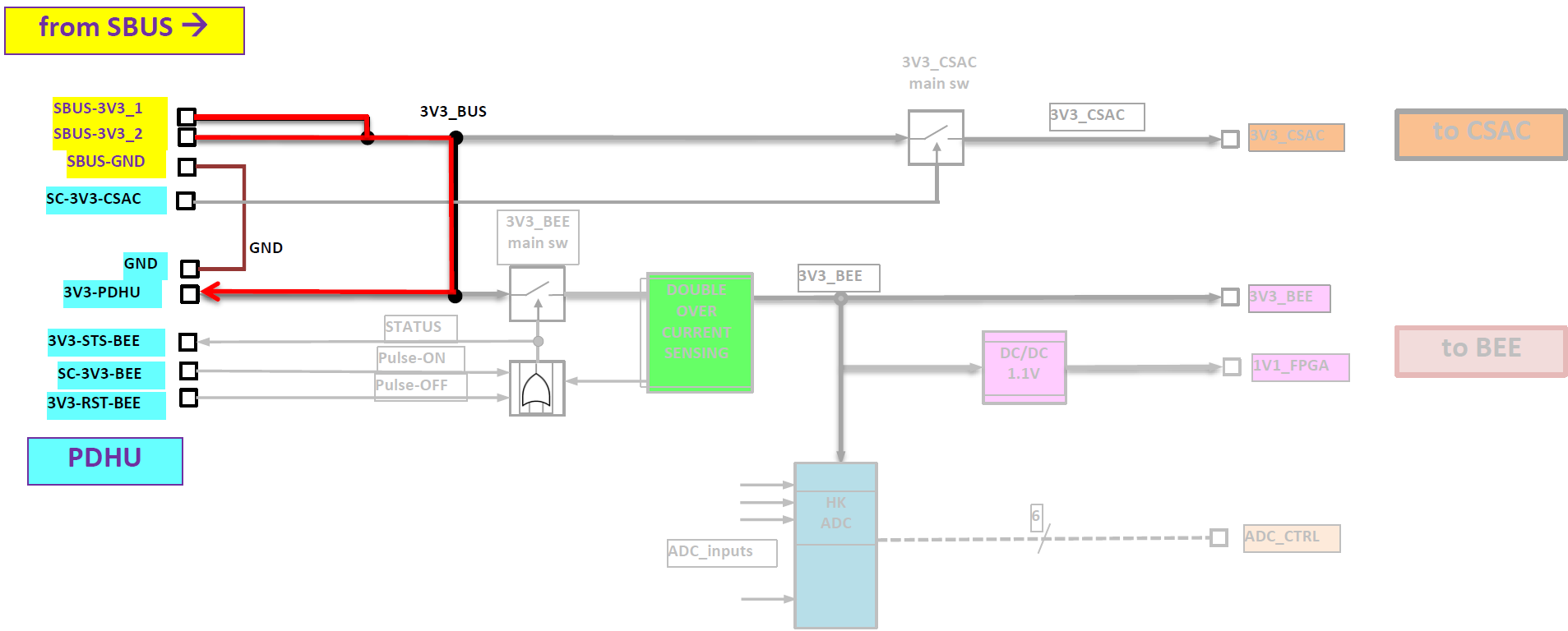} 
\caption{\space}
\label{fig:subim1}
\end{subfigure}
\begin{subfigure}{0.5\textwidth}
\includegraphics[width=0.9\linewidth, height=4cm]{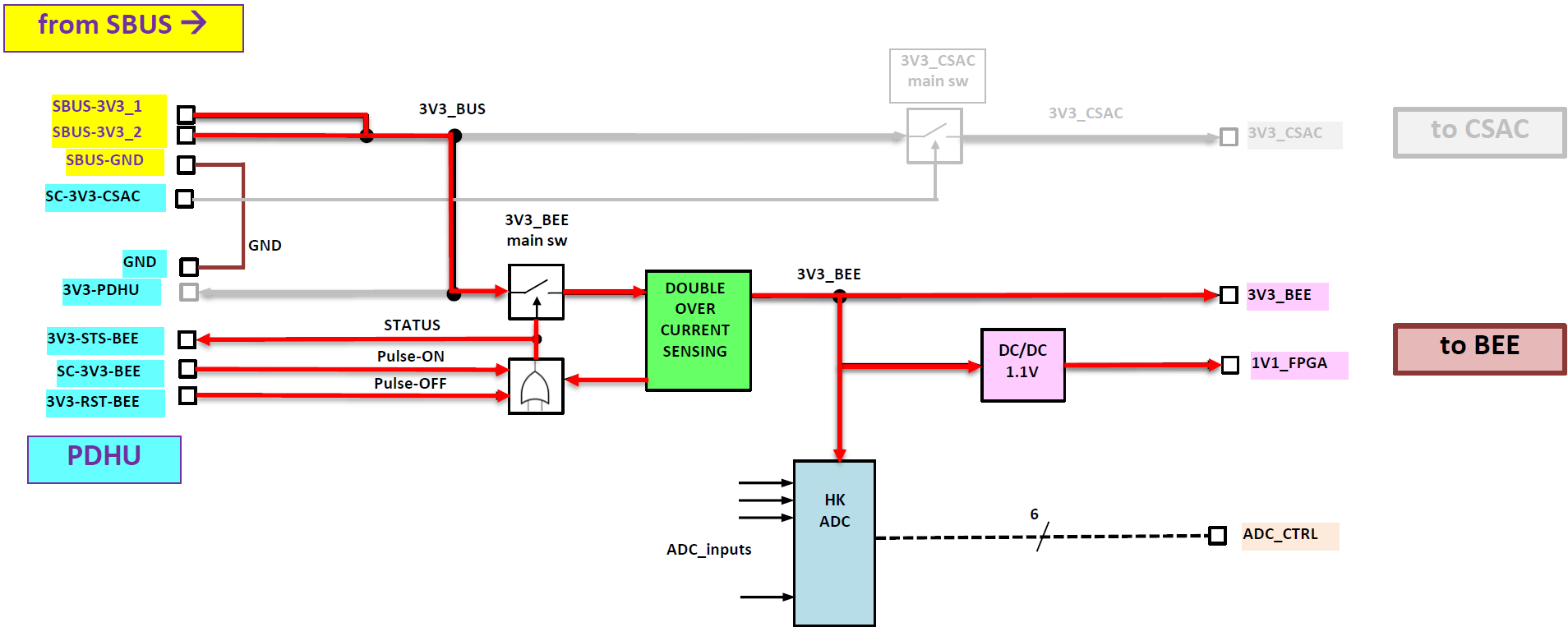}
\caption{\space}
\label{fig:subim6}
\end{subfigure}
\begin{center}
\begin{subfigure}{0.5\textwidth}
\includegraphics[width=0.9\linewidth, height=4cm]{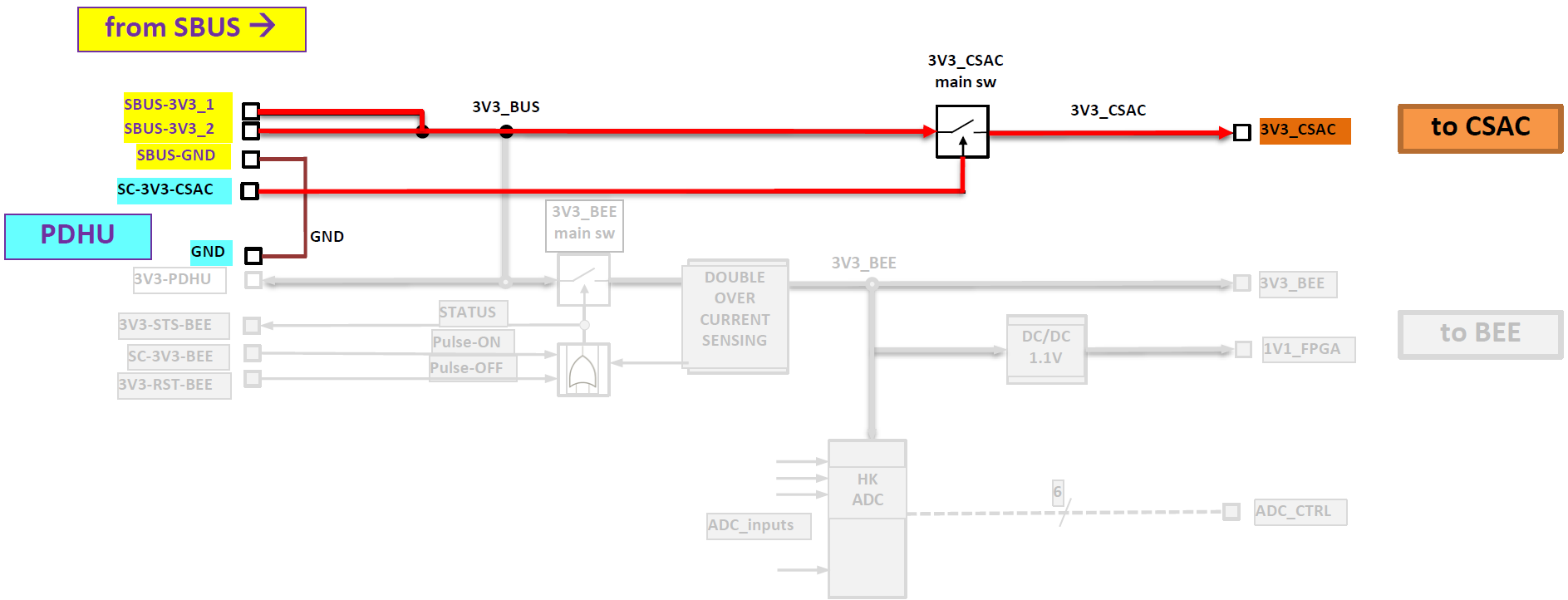} 
\caption{\space}
\label{fig:subim7}
\end{subfigure}
\end{center}
\caption{Block diagram representation of HERMES-PSU 3V3(a)PDHU, (b)BEE, (c)CSAC section}
\label{fig:figu6}
\end{figure}

The 3V3 voltage is provided to the PSU via the SBUS-3V3 and supplies the various parts of the payload, in particular the PDHU, the BEE and the CSAC. In Figure \ref{fig:figu5} we show the block diagram of the PSU in which the 3V3 voltages are distributed to the various sections.

In Figure \ref{fig:figu6} the different section of the PSU supplied by the 3V3 voltage are highlighted.  On the PSU there is a line that is neither monitored nor controlled, while it provides the 3V3 voltage to the PDHU. Once the PDHU is switched on, it can control the closure of the switch that enables the BEE to be powered (Fig. \ref{fig:subim1}). The 3V3-BEE is the main power supply line for the entire PSU and it is highlighted in Figure \ref{fig:subim6}.
The switch that enables the BEE to be powered on is controlled by the PDHU and it was designed as a latch. Thus, if the PDHU shuts down, the BEE remains switched on. The 3V3-BEE subsection generates also the 1V1 voltage which is required to supply power to the FPGA banks. The 3V3-BEE also powers the ADC from which the HKs signals are generated.

Since HERMES will consist of a swarm of CubeSATs, it is necessary to synchronise the acquisitions made by any individual nanosatellite.
Figure \ref{fig:subim7} highlights the part that constitutes the 3V3-CSAC, which powers the CSAC (Chip Scale Atomic Clock) dedicated to the synchronisation of the recording of events detected by the different CubeSats.

\subsection{Voltages required by the FEE}
\begin{figure} [h!]
   \begin{center}
   \includegraphics[height=10cm]{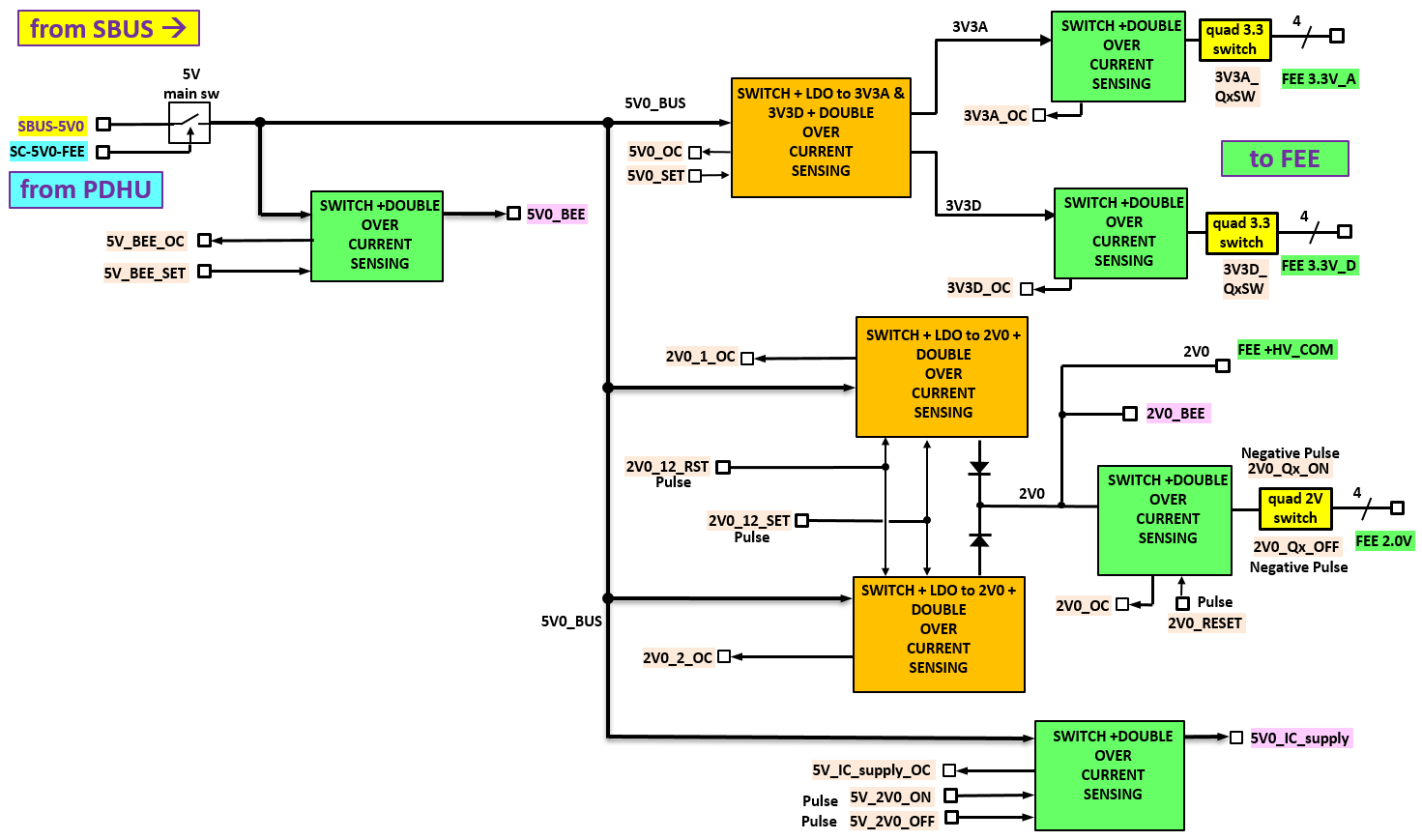}
   \end{center}
   \caption{Block diagram representation of HERMES-PSU 5V0 section.}
  \label{fig:figu7} 
  \end{figure}

The FEE of the HERMES hosts the LYRA ASICs, which require several voltages to properly work. Designing the power supply of a mixed signal device as LYRA, often some doubts arise, whether to use a single power supply for both analog and digital sections or sepated power supplies. The supply stage of the FEE has been carefully designed in order to not degrade the high performances of the LYRAs, so the architecture with separated supplies has been chosen.  To follow the architecture description, reference should be made to Figure 3 because most PSU sections have this structure (BUS switch + LINE switch). Figure \ref{fig:figu7} shows the block diagram of the distribution of the SBUS-5V0 voltage to the various subsections.

   \begin{figure} [h!]
   \begin{center}
   \includegraphics[height=2cm]{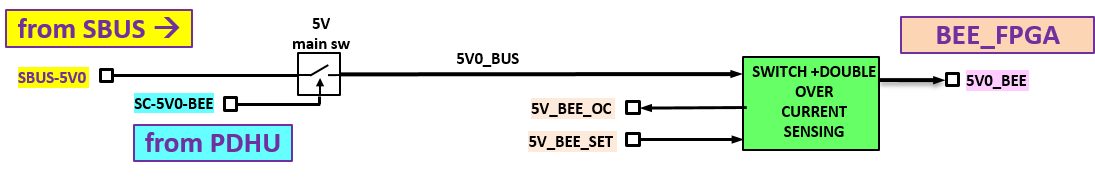}
    \end{center}
  \label{fig:5V0_FEE-BEE} 
  \end{figure} 
   \begin{figure} [h!]
   \begin{center}
   \includegraphics[height=4cm]{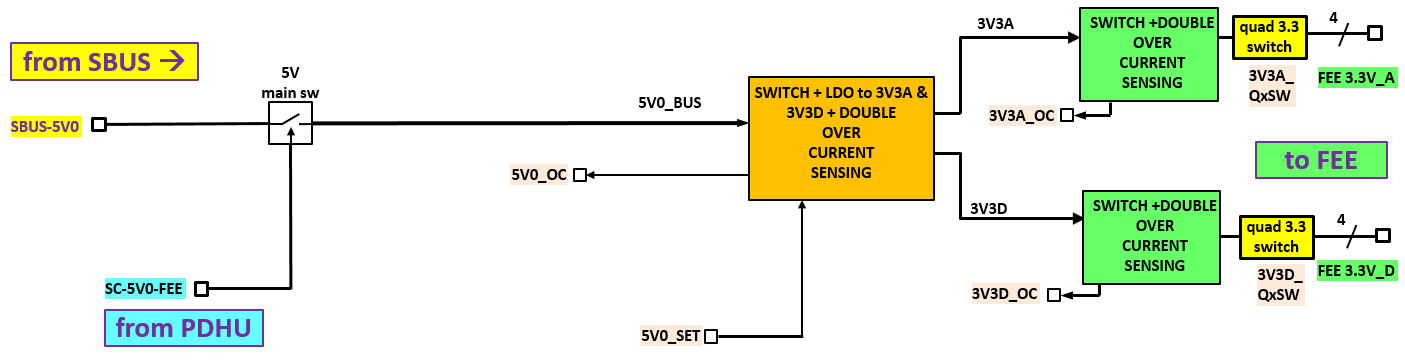}
   \end{center}
   \caption {Block diagram representation of HERMES-PSU 5V0-FEE or 5V0-BEE section (up), HERMES-PSU voltage generation of the FEE (bottom). The thresholds of the ’Overcurrent Sensing’ blocks are gradually decreasing from left to right.}
  \label{fig:image8} 
  \end{figure}

The first part of 5V0-FEE circuitry is identical to 5V0-BEE and the block diagram valid for the description of the 5V0-BEE and 5V-FEE sections is therefore given in the upper section of Figure \ref{fig:image8}.

The nomenclature 5V0-FEE indicates the 5V section from which the voltages necessary for the proper functioning of the FEE are generated. Specifically, the different voltages required are, respectively: 
\begin{itemize}
    \item\textbf{3V3A} required to power the analogue section of the FEE
     \item\textbf{3V3D} required to supply the digital section of the FEE
\end{itemize}

The voltage rails 3V3A and 3V3D are generated by two Analog Devices LT3042 supplied by 5V0-FEE. The LT3042s have been selected among several LDO devices for the very high value of the Power Supply Rejection Ratio (PSRR) over a large bandwidth. Essentially the PSRR expresses the device capability to reject the input ripple and avoid transmitting it to the output. Then this characteristic is very important when a low noise supply is required but the voltage source is switching as in our case. 
In all the sections in which the load is supplied via quadrant switches, first the quadrant switch must be closed, and then the line switch can be closed. This allows to take advantage of the soft-starts present in the LDOs. In the sections where there are no soft-starts, it is possible to take benefit of the fact that the current monitors are still switched off.

\subsection{2V0 voltage line}
\begin{figure} [h!]
   \begin{center}
   \includegraphics[height=8cm]{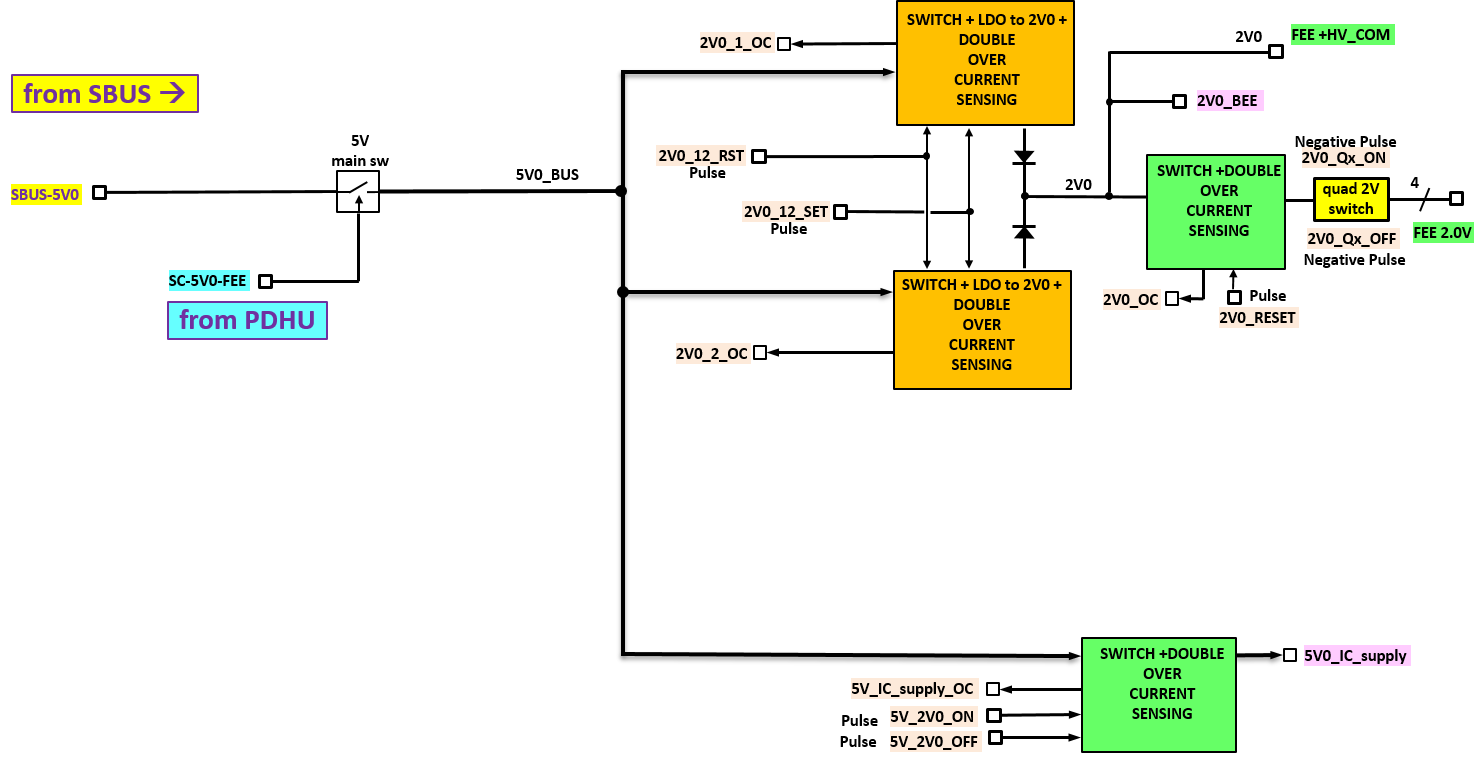}
   \end{center}
   \caption{Block diagram representation of HERMES-PSU 2V0 section. The thresholds of the `Overcurrent Sensing' blocks are gradually decreasing from left to right.}
  \label{fig:figu9} 
  \end{figure}

The line that required the most attention is the line with which 2V0 is generated. This voltage is used to supply the last polarisation ring close to the cathode of the SDD detector. This polarisation prevents damage to LYRAs when the FEE has to be switched off due to latch-up but the HV section is still switched on.
It was therefore decided to divide the 2V0 into two stages:
\begin{itemize}
    \item\textbf{FIRST STAGE}: in which the 2V0 is generated by means of two LDOs arranged on two parallel branches;
     \item\textbf{SECOND STAGE} : in which 2V0 is distributed to the various quadrants and with which the control ICs on the 2V0 line are supplied.
\end{itemize}
The block diagram constituting the generation and control section of the 2V0 is shown in Figure \ref{fig:figu9}.

\subsubsection{FIRST STAGE: Parallel branches}
The two-branch split structure is designed to implement cross-control on the two branches and guarantees the 2V0 to the last polarization ring of the SDD detector. The two branches are identical.

On the individual branch there is a normally closed switch driven by the BEE or by a current monitor. This switch allows to provide the 5V0 voltage to the rest of the branch by a monitored $R_\mathrm{sense}$. Following on from the $R_\mathrm{sense}$ a LDO generates the 2V0 voltage. To ensure redundant control, the current monitor that monitors the $R_\mathrm{sense}$ of the first branch is supplied downstream of the $R_\mathrm{sense}$ of the second branch (see Figure \ref{fig:figu11}).
The two branches are connected to the same node in order to guarantee the presence of the 2V0 downstream in the event that one of the two LDOs is damaged (leaving a single branch open). Downstream of the junction node of the two branches is a pin that provides the 2V0 to the last ring of the SDD detector when the OC condition is detected on the 2V0 quadrants.

\subsubsection{SECOND STAGE: Quadrants section and 2V0 control integrated circuit (IC) supply}
In this section we have the usual structure consisting of a LINE switch and a $R_\mathrm{sense}$ monitored by a current monitor. The difference with the sections seen so far is that the switch is implemented by ICs, which must be monitored since they could be affected by latch-up events. This has led to the idea of a 5V0 line supplying power to the ICs on the 2V0 line, monitored by a redundant control.

%

\subsection{Redundant current monitor}
Different structures for the line current monitor were implemented on the PSU depending on the sections to be supplied: e.g., there are voltages which can be switched off instantaneously and there are voltages which must be switched off after the other sections have been safed. This required to take special care to implement a redundant line current monitor. 
In the first redundant control structure, two current monitors were arranged to monitor a single $R_\mathrm{sense}$. The two current monitors are supplied downstream of the $R_\mathrm{sense}$, in this way the overcurrent sensing blocks are 'blind' to the inrush current during power-up.

\begin{figure} [h!]
   \begin{center}
   \includegraphics[height=6cm]{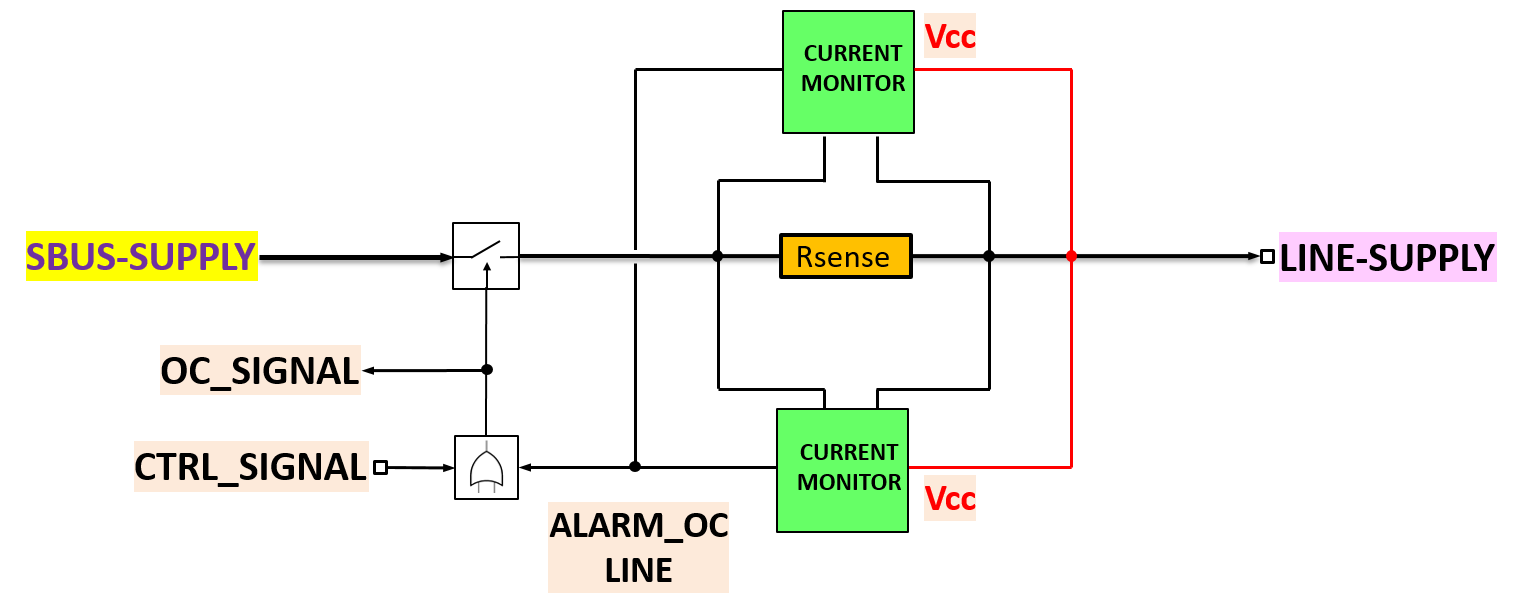}
   \end{center}
   \caption{Block diagram representation of HERMES-PSU double current monitor}
  \label{fig:figu10} 
  \end{figure}

The corresponding block diagram is shown in Figure \ref{fig:figu10}.

As can be seen in the figure, the switch is controlled by both the BEE and the current monitor blocks.

This structure has been used in all those sections where the line voltage can be abruptly removed when an OC condition occurs. In fact, as soon as one of the two current monitor blocks detects the OC condition, the switch is commanded to open and the load is disconnected. In this condition, the BEE must immediately bring the control pin to a low state to avoid to prematurely closing again the switch (e.g, while the OC condition has not yet extinguished).

\begin{figure} [h!]
   \begin{center}
   \includegraphics[height=8cm]{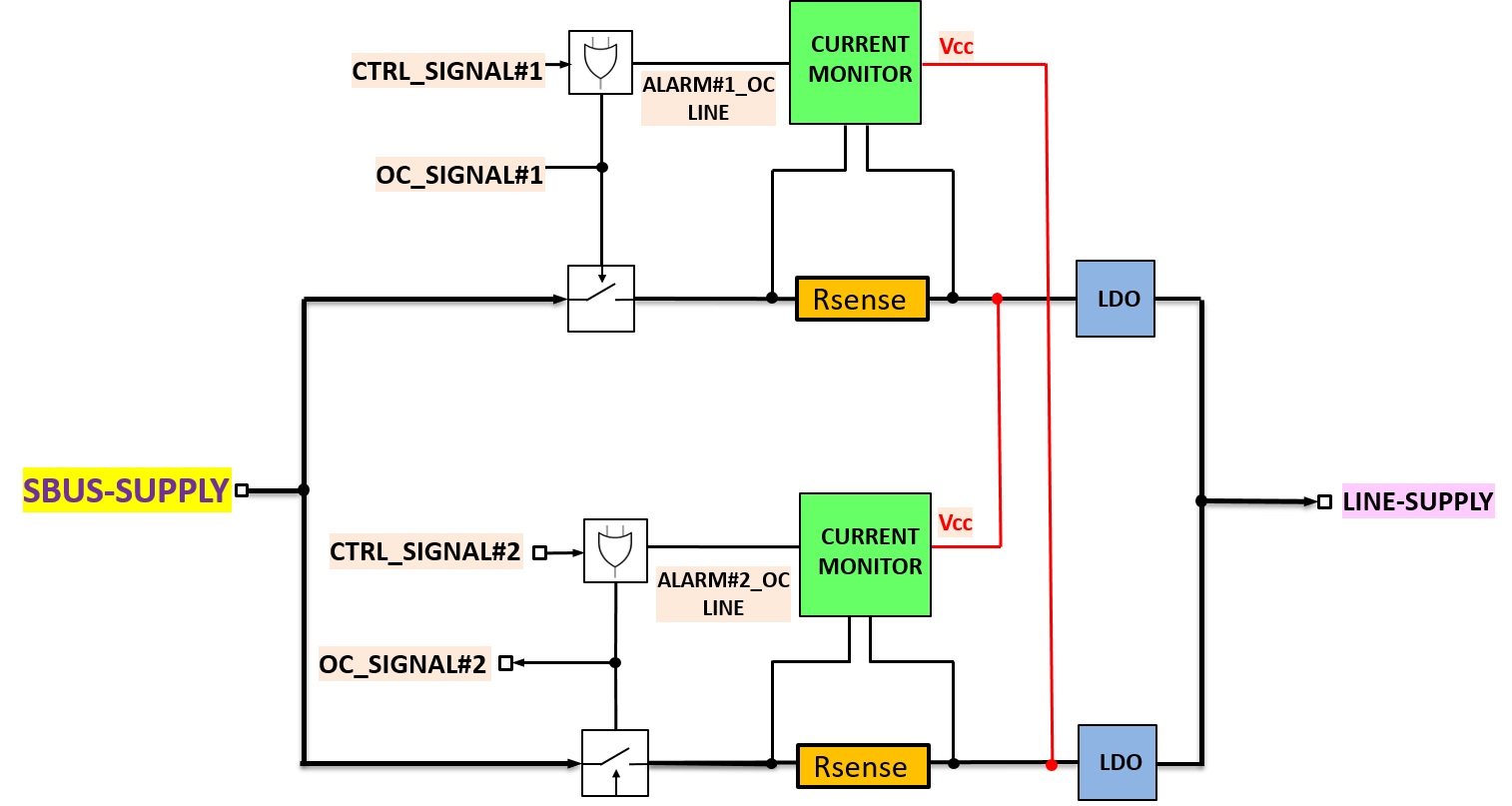}
   \end{center}
   \caption{Block diagram representation of HERMES-PSU split line cross current monitor}
  \label{fig:figu11} 
  \end{figure}

The second redundant control structure was implemented by splitting the line voltage into two parallel lines. On each line, a switch and a current monitor were placed. The corresponding block diagram is shown in Figure \ref{fig:figu11}.

As can be seen in the figure, the current monitor which monitors the $R_\mathrm{sense}$ of one branch is supplied downstream of the $R_\mathrm{sense}$ of the parallel branch. In this way, if one of the two current monitors reveals an OC condition, it opens the switch of the branch to prevent damage to the components.

This allows the branch in which an OC has occurred to be safe-guarded, but at the same time allows the load to continue to be supplied with the nominal voltage and also allows a specific switch-off procedure to be applied. Note that during the shutdown procedure, the current monitor block of the branch that has been closed will be switched off, and therefore during this phase it will not be possible to monitor the current of this branch. However, the duration of the shutdown procedure is commensurate with the risk of this failed current monitoring.

A peculiarity of some switches used with this structure is that they remain in their state (open/closed) even in the absence of control by the BEE. This makes it possible to safeguard the system by means of the current monitors, but at the same time ensures that a correct switch-off procedure is carried out even without direct control.

\subsection{High-Voltage (HV) section}
The HV section is required to generate the polarisation voltage of the SDD detector. The HV voltage is generated from the 12V0 rail supplied by the satellite BUS (SBUS-12V0). Thus, the HV section can be divided into two parts:
\begin{itemize}
    \item \textbf{a low-voltage (LV) section} in which the voltage to be supplied to the DC-DC is managed by controlling a series of switches;
    \item \textbf{a hybrid section (HV-LV)} in which the first one contains the DC-DC which generates the HV, and the second one contains both the feedback line to obtain the HK value and the section which generates the control ramp for the HV.
\end{itemize}

The first section consists of two switches in series: the first is controlled only by the PDHU and the second is controlled by the BEE and the current monitor. The first switch has been designed to open if control by the PDHU should fail, the second switch mantains its state (open/closed) should control by the BEE fail.

The second section consists of a block containing the DC-DC which generates the HV and the network which implement the control ramp, and another block consisting of the HV housekeepings generation network.

HV is generated by a ramp that lasts 6 seconds. To generate the HV with a ramp, a network consisting of discrete components was designed in order to prevent that this section from being affected by latch-ups. This network was connected to the control pins of the DC-DC converter (Pico 12SAR250) and has been designed in such a way that if the BEE control should fail then a ramp-down will be generated.

\begin{figure} [h!]
   \begin{center}
   \includegraphics[height=9cm]{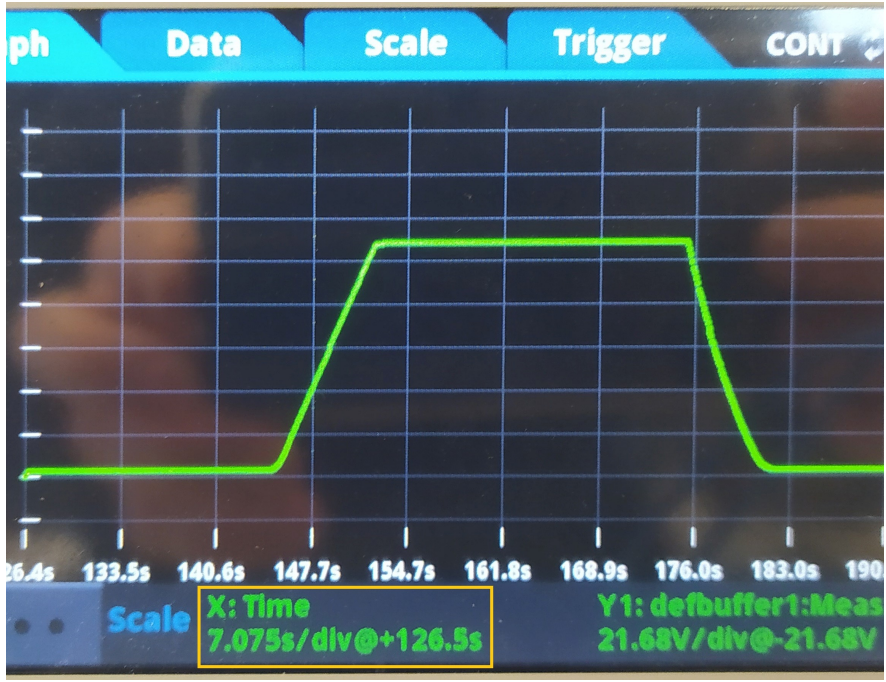}
   \end{center}
   \caption{Ramp-HV curve}
  \label{fig:figu12} 
  \end{figure}

The polarisation voltage of the SDDs is a negative voltage and since only positive voltages are present on the BUS, an inverter is needed to obtain the HK signal of the HV. This inverter is used to generate the supply voltage for the buffers used to read the HV value. The HK signal is buffered twice in order to not alter the value to be sampled by the ADC (obtained by a resistive divider) and to eliminate any offset between GND-HV and GND-LV.
Another section of the HV is split in the individual FEE quadrants. The control section of the HV dial switches is powered by 5V, which is supplied by an LV switch controlled by the BEE. This LV switch has been implemented with a structure that maintains the open/closed state should the BEE switch off.

\subsection{Overcurrent Sensing performance}
To implement the Overcurrent (OC) Sensing block, an integrated circuit was selected, based on the requirement of obtaining as much information and functionality as possible with less external components as possible. Additional requirements were also addressed in order to keep an analog signal proportional to the line current to be monitored. When this signal exceeds an externally settable threshold voltage, then the OC condition occurs. So that, the integrated component has been equipped of a pin which flags the occurrence of the OC condition. Another required function of the circuit is the latching of the OC condition.

This integrated circuit in orbit could be itself affected by latch-up and thus damaged \cite{Peter1}. Considering that the probability of such an event occurrence is non-zero, a redundant use of the integrated circuit was considered. In particular, one of the structures seen in the previous section will be adopted depending on the section to be monitored. 
Regardless of the implemented structure, the chosen IC ensures that as soon as an OC is detected, its ALERT pin goes to logic low level, the line opens and the corresponding load is switched off.

The integrated circuit INA301-Q1 was tested to experimentally determine the response times for different values of the load capacity.  Initially, in the test setup a resistive load has been fixed in order to have a nominal operating current, and a trigger threshold has been set accordingly. In parallel to the fixed load, a smaller resistor is connected by means of a push-button, thus inducing the OC condition ($V_\mathrm{LINE} > V_\mathrm{THRESHOLD}$) and the integrated circuit will bring the ALERT line to a logic low level. The circuit made to induce the OC condition and an acquisition made, with the corresponding oscilloscope channel settings, are shown in sections (a) and (b) of Figure \ref{fig:OC_detect}, respectively.

\begin{figure} [h] 
\begin{subfigure}{0.6\textwidth}
\includegraphics[width=0.7\linewidth, height=3cm]{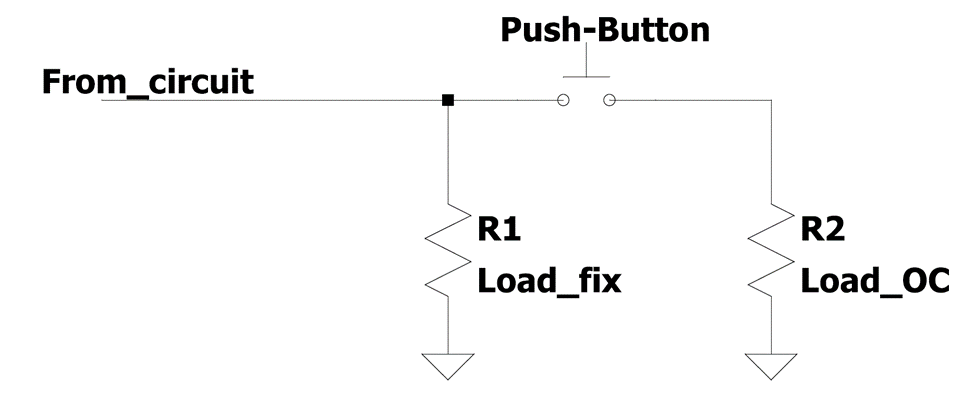}
\caption{\space}
\label{fig:Load_R}
\end{subfigure}
\begin{subfigure}{0.5\textwidth}
\includegraphics[width=0.8\linewidth, height=3cm]{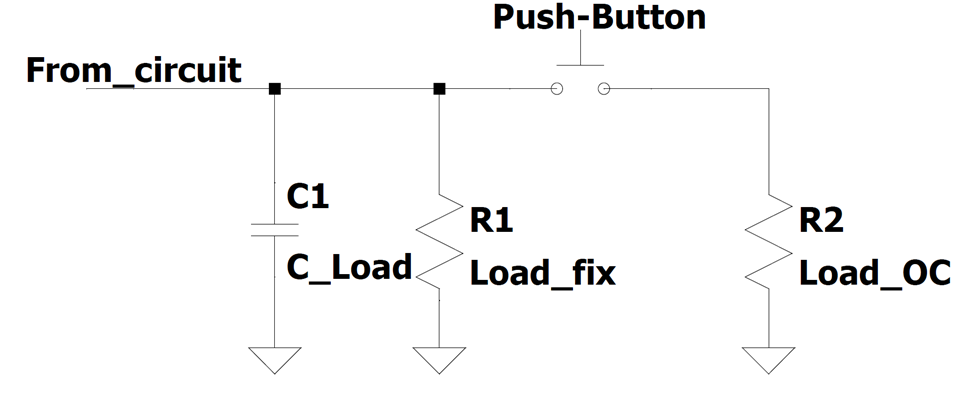}
\label{fig:OC_detect}
\end{subfigure}
\begin{subfigure}{0.8\textwidth}
\begin{center}
\includegraphics[width=0.9\linewidth, height=6cm]{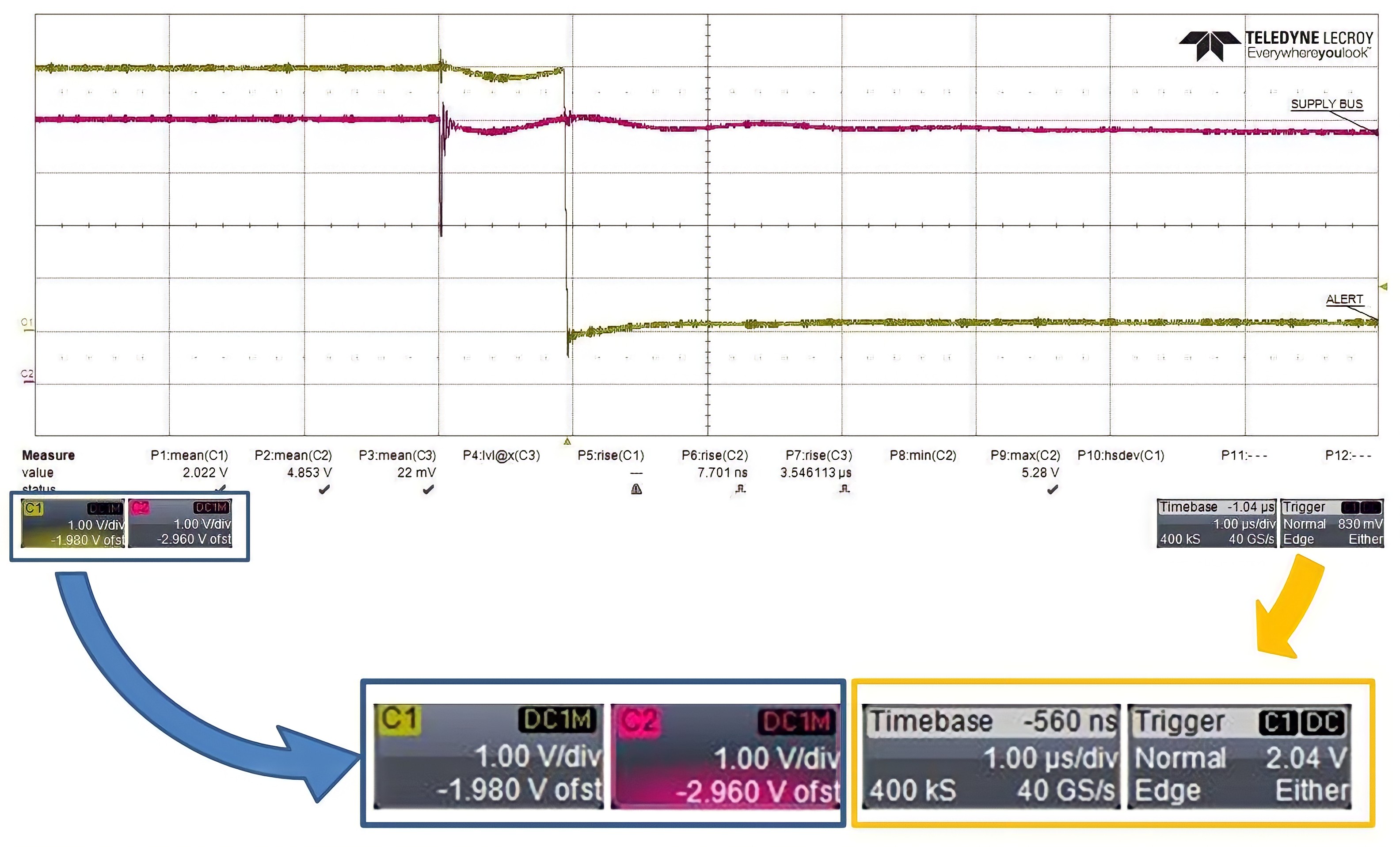}
\caption{\space}
\label{fig:OC_detect}
\end{center}
\end{subfigure}
\caption{Circuit structure to induce the OC condition(a) and an acquisition of OC-detection(b)}
\label{fig:OC_detect}
\end{figure}
   
Specifically, the line voltage (in violet) and the voltage at the ALERT pin (in yellow) were acquired. The line voltage was monitored to see its variations when a button is pressed.
A capacitor was then added in parallel to R1 to assess how much the circuit slows down in detecting the OC condition. The test was repeated for three values of $C_\mathrm{LOAD}$, 1~µF, 10~µF and 100~µF respectively, to see the corresponding variations in response time.

\begin{figure} [h!] 
\includegraphics[width=1\linewidth, height=9cm]{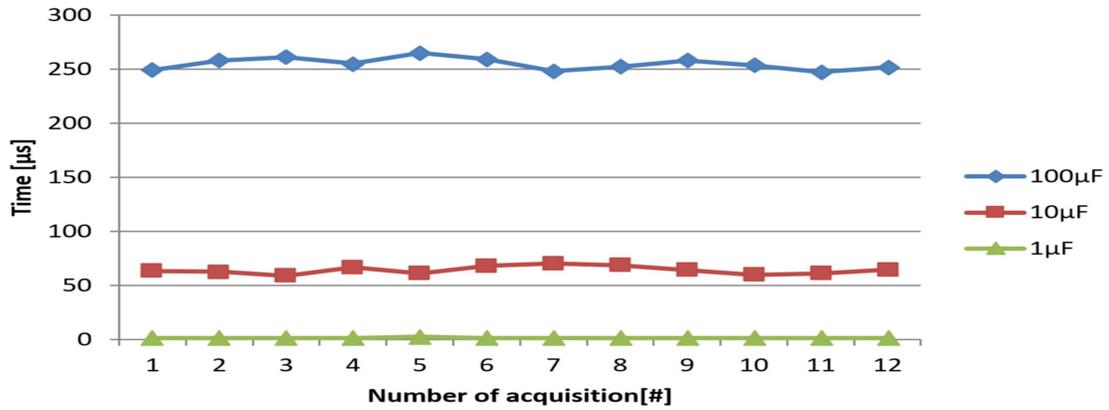}
\caption{Variation in OC detection time when varying load capacity}
\label{fig:fig14}
\end{figure}

The trend of the drift time and the setting of the oscilloscope channels are shown in Figure \ref{fig:fig14} and Figure \ref{fig:trendOC_detect}.  
   
\begin{figure} [h!] 
\begin{subfigure}{0.6\textwidth}
\includegraphics[width=1.1\linewidth, height=9cm]{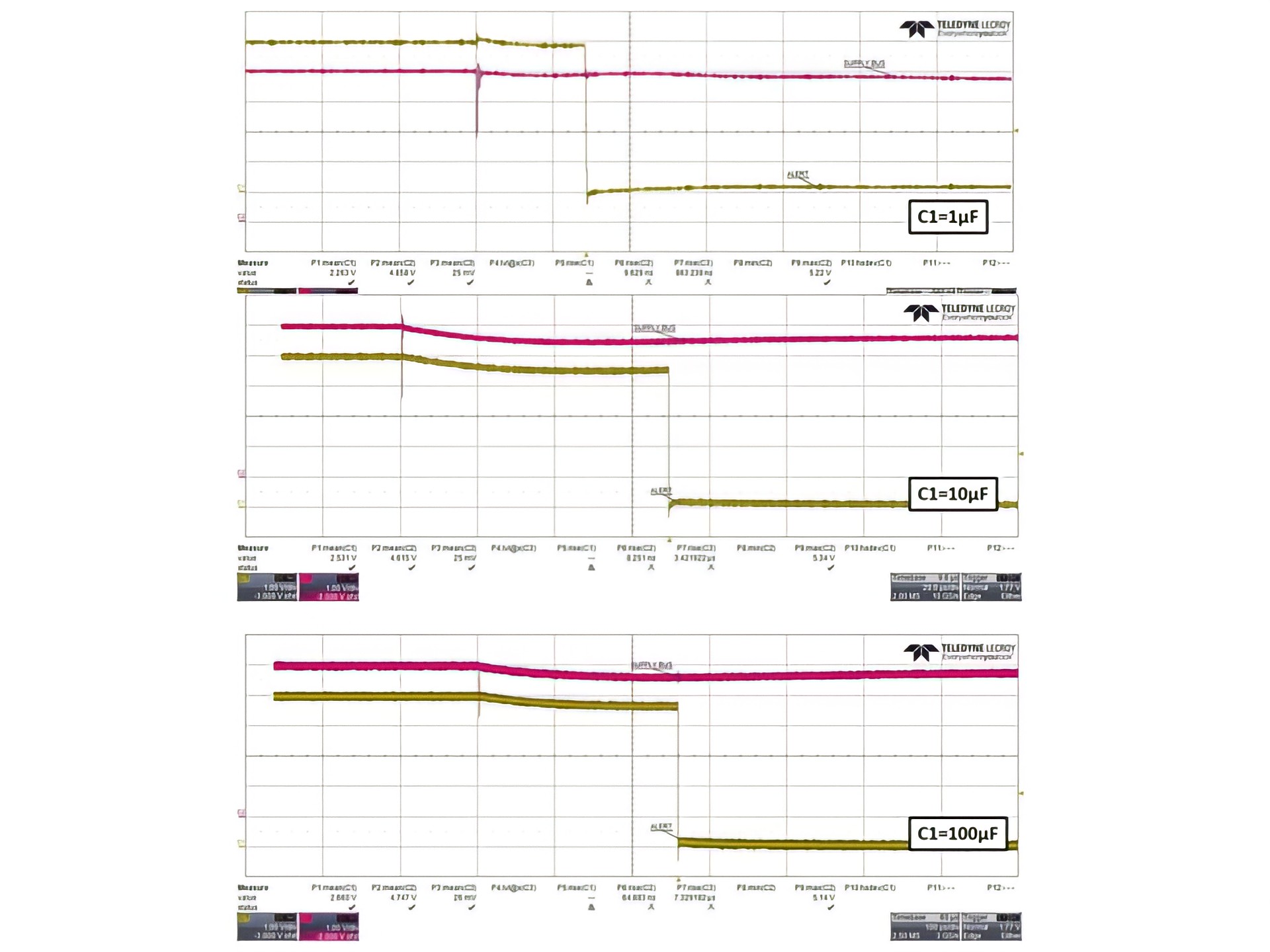}
\caption{\space}
\label{fig:driftTimeDetect}
\end{subfigure}
\begin{subfigure}{0.5\textwidth}
\includegraphics[width=0.8\linewidth, height=6cm]{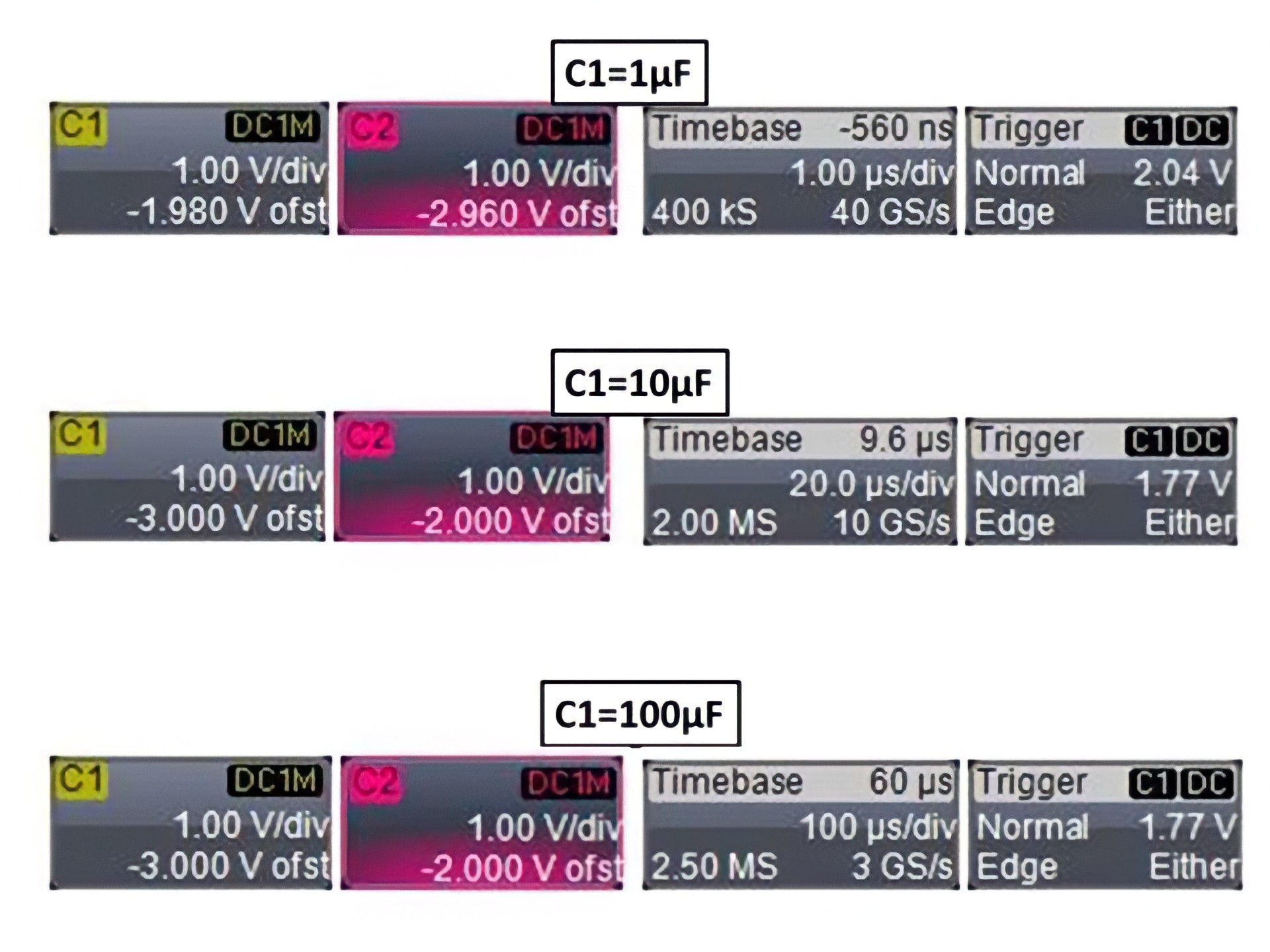}
\caption{\space}
\label{fig:trendOC_detect}
\end{subfigure}
\caption{Variation in OC detection time when varying load capacity (a) and oscilloscope acquisition of OC detection for different load capacitance values (b)}
\label{fig:trendOC_detect}
\end{figure}


\subsection{Housekeeping Analogue to Digital Converter (ADC)}
To check that the PSU is correctly feeding the different sections, both voltages and currents are monitored at strategic points. Specifically, the housekeeping signals are acquired using an analogue-to-digital converter (ADC). The ADC chosen is the MAX11643, an 8-bit ADC, 16 channels, internal reference, a serial FIFO memory and a maximum sampling rate of 300~ksps. The sampling frequency used for acquisitions can vary in the range 15~mHz--4~Hz. In order to sample the voltages of the various sections, in some cases they had to be adapted appropriately to the operating range of the ADC. The ADC can communicate with the BEE via an SPI interface at a maximum frequency of 10~MHz.

\subsection{Radiation Test of COTS component}
As mentioned above, commercial components were used to build the PSU. Although particular designs were implemented to protect the electronics from possible latch-up events \cite{Peter1}, a number of tests were carried out to find out how much the components could be exposed to radiation particles such as protons before being permanently damaged. To assess this, it was decided to irradiate the components considered critical for their relevance to the proper functioning of the payload with protons. It has been decided to carry out these tests at the proton synchrotron accelerator of the Protontherapy facility in Trento, Italy, managed by TIFPA  (Trento Institute for Fundamental Physics and Applications). The components irradiated were: 
\begin{itemize}
    \item 3 DC-DC PICO 12SAR250 to generate the HV,
    \item 2 DC-DCs LMZ30602 to generate the 1V1,
    \item 3 ADAQs, ADAQ7980 i.e. ADCs sampling the signals from the FEE,
    \item 2 ADCs MAX11643 sampling the signals from HKs
\end{itemize}
Among these components, only one of the three DC-DC was found to be permanently damaged, albeit at an equivalent dose to 2 years in an equatorial LEO.
A more in-depth discussion of these results will be presented in a further paper.

\section{Conclusion}
On the HERMES flight PSU boards produced so far, many tests were carried out in order to assess the correct functioning under different working conditions. Some tests were repeated in a climatic chamber (temperature range -10 -- +10 °C) in order to assess the thermal drift of certain components, such as the HV DC-DC converter.

The PSU board was intensively tested both on its own and integrated with the payload. In some cases, these tests were performed with customised setups in order to test the individual PSU line section. These setups allowed us to highlight a number of criticalities that were fixed in order to guarantee greater reliability of the circuit.

During the FM (Flight Module) integration campaign the entire integrated payload has been extensively tested.
Initial difficulties encountered when interfacing PSU-FEE and PSU-BEE  were overcome by adjusting the setup and fine-tuning the power-up procedure, minimising unwanted/parasitic effects compromising the correct power-up of the various sections of the PSU.

\acknowledgments{
This work has been carried out in the framework of the HERMES-TP and HERMES-SP collaborations. We acknowledge support from the European Union Horizon 2020 Research and Innovation Framework Programme under grant agreement HERMES-Scientific Pathfinder n. 821896 and from ASI-INAF Accordo Attuativo HERMES Technologic Pathfinder n. 2018-10-H.1-2020.}

\bibliography{report} 

\begin{thebibliography}{1}

\bibitem{Fiore2}
Fiore, F. et~al., ``{The HERMES-technologic and scientific pathfinder},'' {\em
  Proceedings of SPIE}~{\bf 11444-166} (2020).

\bibitem{Sanna1}
Sanna, A. et~al., ``{Timing techniques applied to distributed modular
  high-energy astronomy: the H.E.R.M.E.S. project},'' {\em Proceedings of
  SPIE}~{\bf 11444-251} (2020).

\bibitem{Burd3}
Burderi, L. et~al., ``{GrailQuest \& HERMES: Hunting for Gravitational Wave
  Electromagnetic Counterparts and Probing Space-Time Quantum Foam},'' {\em
  Proceedings of SPIE}~{\bf 11444-252} (2020).

\bibitem{Evang4}
Evangelista, Y. et~al., ``{The scientific payload on-board the HERMES-TP and
  HERMES-SP CubeSat missions},'' {\em Proceedings of SPIE}~{\bf 11444-168}
  (2020).

\bibitem{Fab-LDO}
Fabio P. Lo~Gerfo, G.~S., ``{Misura del PSRR nei regolatori di tensione ad alte
  prestazioni},'' {\em INAF Technical Reports - Rapporti Tecnici INAF} (127)
  (2022).

\bibitem{Peter1}
Petersen, E., ``{The SEU figure of merit and proton upset rate calculations},''
  {\em IEEE Transactions on Nuclear Science}~{\bf 45}(6),  2550--2562 (1998).

\end{thebibliography}
\bibliographystyle{spiebib} 

\end{document}